\let\orilabel\label
\documentclass[%
 reprint,
 amsmath,amssymb,
 aps,
prb,
]{revtex4-2}

\usepackage{graphicx}% Include figure files
\usepackage{dcolumn}% Align table columns on decimal point
\usepackage{bm}% bold math
\usepackage{xcolor}%to choose color parts
\usepackage{soul}%to highlight parts
\usepackage{setspace}
\let\label\orilabel
\usepackage{hyperref}% add hypertext capabilities

\allowdisplaybreaks

\begin{document}
%\setstretch{1.3}

\title{Depletion-mode N-polar AlN-based high electron mobility transistors with improved on/off ratios}

\author{Xu Yang}
 \email{yangxu2025@ime.ac.cn}
\affiliation{Center for Integrated Research of Future Electronics, Institute of Materials and Systems for Sustainability, Nagoya University, Furo-Cho, Chikusa-Ku, Nagoya, 464-8601 Japan}%
\affiliation{Institute of Microelectronics, Chinese Academy of Sciences, Beijing 100029, China}
\author{Sheng Zhang}
\affiliation{Institute of Microelectronics, Chinese Academy of Sciences, Beijing 100029, China}
\author{Itsuki Furuhashi}
\affiliation{Center for Integrated Research of Future Electronics, Institute of Materials and Systems for Sustainability, Nagoya University, Furo-Cho, Chikusa-Ku, Nagoya, 464-8601 Japan}%
\author{Ke Wei}
\author{Xinhua Wang}
 \email{wangxinhua@ime.ac.cn}
\author{Xinyu Liu}
\affiliation{Institute of Microelectronics, Chinese Academy of Sciences, Beijing 100029, China}
\author{Markus Pristovsek}
 \email{markus.pristovsek@nagoya-u.jp}
\affiliation{Center for Integrated Research of Future Electronics, Institute of Materials and Systems for Sustainability, Nagoya University, Furo-Cho, Chikusa-Ku, Nagoya, 464-8601 Japan}%

\date{\today}% It is always \today, today,

\begin{abstract}
We report N-polar AlN-based high-electron mobility transistors (HEMTs) with a GaN channel thickness of 5.2\,nm on N-polar AlN on sapphire. The threshold voltage is around -2.4 to -3.0\,V with saturation currents over 240\,mA/mm and on/off ratios as high as $10^{4}$, much higher than previously reported N-polar AlN-based HEMTs. The high on/off ratio is attributed to the use of an abrupt AlN/GaN heterostructure with a dedicated AlN transition layer, together with improved gate leakage. The high frequency properties as well as the on-resistance of $\approx 20\,\Omega\,$mm are all limited by the 2000\,$\Omega/\Box$ sheet resistance of the channel layer.
\end{abstract}

\maketitle

\section{Introduction}

N-polar AlN-based high electron mobility transistors (HEMTs) have intrinsic advantages compared to the more mainstream N-polar GaN-based HEMTs and the commercialized Ga-polar HEMTs. AlN is an excellent material for the buffer layer. First, the ultrawide bandgap of AlN results in a very high breakdown voltage and it is highly insulating, suppressing buffer leakage effectively, which can be a large problem for N-polar GaN-based HEMTs. Furthermore, AlN has a high thermal conductivity, higher than GaN or sapphire, allowing for more heat extraction and operate the devices at higher power levels. The very high contrast in polarisation also leads to very high sheet carrier densities at the GaN/AlN heterointerface, close to $5\times 10^{13}$\,cm$^{-2}$. Finally, the large difference in the bandgap between AlN and GaN results in highly confined channels that negate short-channel effects and can reduce current collapse \cite{Zhang25Bristol}.

So far, the progress has been limited by the growth of N-polar AlN itself. Only Cornell University \cite{Singhal22,Kim22hemt,Kim23hemt}, Yamaguchi University \cite{Isono20,Miyamoto23}, and us \cite{Pampili24,Pri24,Robin26} have reported smooth enough N-polar AlN to measure a 2-dimensional electron gas (2DEG) when putting a strained GaN channel on top of it. Cornell used bulk AlN substrates that show atomic steps, despite their low misorientation of less than 0.3°.  In comparison, hetero-epitaxial growth of AlN usually reports rough surfaces with many hillocks (e.g.\ \cite{Hu23AlN,Namikawa23,Yang25SiCepi}) until misorientation angles of 2° and larger are used \cite{Isono20,Pampili24,Namikawa23} together with extreme Al-rich growth conditions and high temperatures. Otherwise, the surface is dominated by hexagonal hillocks with sizes depending on surface diffusion. The extreme growth conditions needed for hetero-epitaxy on sapphire lead to much higher oxygen incorporation by metal-organic vapor phase epitaxy (MOVPE) exceeding $10^{19}$\,cm$^{-3}$ \cite{Pri24,Kowaki24}, while lower oxygen levels were reported for molecular beam epitaxy (MBE) on bulk N-polar AlN substrate (approximately $8\times 10^{17}$\,cm$^{-3}$) \cite{Singhal22}. Similar levels have been reported for the GaN channel \cite{Miyamoto23}, which is another challenge. Therefore, N-polar AlN-based HEMTs with GaN channels are a very recent field. Therefore, N-polar AlN-based HEMTs with GaN channels are a very recent field. Apart from early studies on N-polar metal–semiconductor field-effect transistors of Si-doped AlN channel \cite{Lemettinen18sims} and polarization-doped AlGaN channel \cite{Lemettinen19}, the first N-polar AlN-based HEMTs with a 2DEG have been reported in 2022 by Cornell University (USA) using MBE on bulk N-polar AlN substrates \cite{Kim22hemt}. After improvements and with scaling, these HEMTs showed very high saturation currents ($I_{sat}$) of over 2500\,mA/mm and good RF properties but with a high threshold voltage below -7\,V \cite{Kim23hemt}. Moreover, these HEMTs did not fully turn off and showed leakage currents of over 5-33\,\% of $I_{sat}$ \cite{Kim22hemt,Kim23hemt}, i.e., low on/off ratios. Slowly thereafter, N-polar AlN-based HEMTs have also been reported by Yamaguchi University (Japan), which were grown on vicinal sapphire substrates by MOVPE \cite{Inahara23,Kowaki24,Zazuli24sims,Zazuli24thick}. The $I_{sat}$ were much lower, increasing from 2-12\,mA/mm for an AlGaN channel HEMT \cite{Inahara23,Miyamoto23} to 243\,mA/mm for the best GaN channel HEMT \cite{Kowaki24}. These devices also showed considerable leakage in off-state, again about 5\,\% of $I_{sat}$ with better threshold voltages of below -3\,V. All the HEMTs from both groups have an AlGaN transition layer between the AlN buffer and the GaN channel. 

Simulations have shown that AlGaN could potentially form a second 2DEG at the AlN/AlGaN interface \cite{Kim23hemt}. Thus, carriers could leak to this layer in the off-state since it would need very high voltages to deplete the buried second 2DEG. Moreover, the charges of the 2DEG in the GaN channel are reduced by the bottom 2DEG, and the AlGaN back barrier would cause additional alloy scattering since part of the 2DEG’s wave function penetrates the AlGaN back barrier. Both effects become more severe with lower Al contents.

Therefore, we decided to focus on N-polar AlN/GaN hetero-interfaces using AlN transition layers as discussed in \cite{Pri24,Robin26} and processed these into HEMTs. In this study, we report on both DC and RF characteristics of N-polar AlN-based HEMTs with a GaN channel grown on sapphire substrates.

\section{Experimental}

\begin{figure}
\includegraphics[width=1.0\columnwidth]{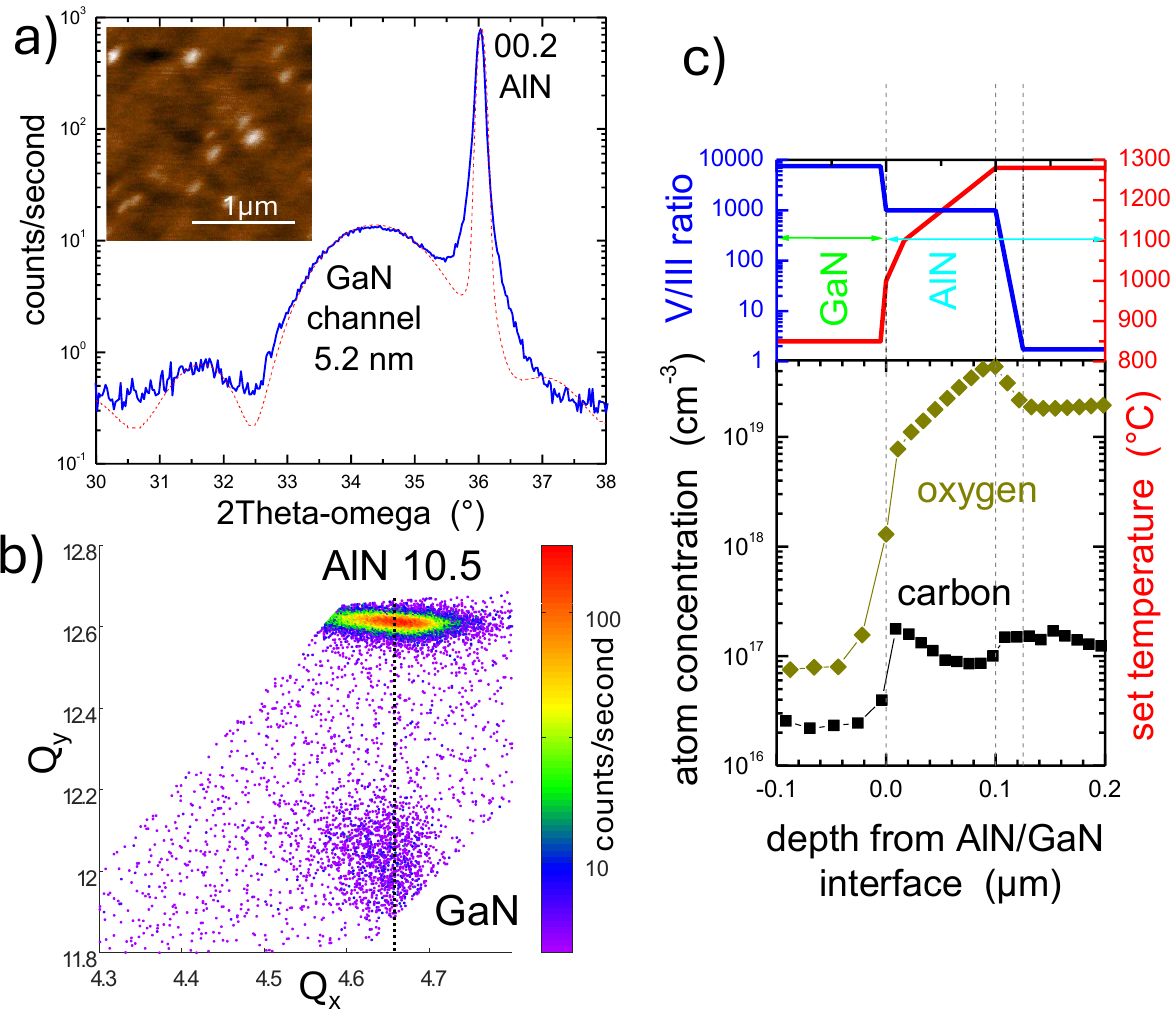}
\caption{a) XRD $2\Theta-\omega$ measurement around 0002 AlN (blue) and simulation (red dashed) of coherent 5.2\,nm GaN (red dashed). Inset: Surface morphology of the GaN channel, exhibiting the RMS surface roughness of 0.22\,nm, scale bar: 1\,$\mu$m. (b) XRD reciprocal space map around the AlN 10\=15 reflection, confirming a fully strained GaN channel without any AlGaN intermixing. (c) SIMS of a comparable sample with double AlN transition layer growth times (for higher resolution). The top panel shows the variation in NH$_3$ flow as V/III ratio and growth temperature.}
\label{i:sample}
\end{figure}

The samples in this study were grown in a 3$\times$2{"} close-coupled showerhead MOVPE reactor from EpiQuest. The substrates were (0001) sapphire 4° misoriented to [11\=20]. The growth details of the N-polar AlN and AlN/GaN heterostructures in this reactor have been reported before \cite{Pampili24,Pri24}. It consists of a nitridation step, an AlN buffer growth step, an AlN transition layer (about 5 nm) where ramping NH$_3$ and temperature from 1300°C to 1000°C and from a V/III ratio of 1.75 to 20000), and after the shortest possible stabilisation at a setpoint of 825°C. The low temperature was chosen to reduce carbon \cite{Pri24} and to get the smoothest surface without step-bunching, similar to literature \cite{Miyamoto23}. Moreover, for Al-polar AlN it was also reported that thicker strained GaN is possible at lower temperatures \cite{Yoshikawa24}, and theoretically lower dislocation density at lower temperatures as well \cite{Yang25dislocations}. For the sample in this study, we used a thin AlN buffer layer of only 111\,nm with a GaN channel thickness of 5.2\,nm. As shown in Fig.~\ref{i:sample}, the samples were characterized with high-resolution X-ray diffraction (XRD), X-ray reflection, atomic force microscopy (AFM) and contactless sheet resistivity via eddy currents. The sheet resistance in the center of about 2000\,$\Omega/\Box$ matches a temperature-dependent Hall effect measurements of a comparable sample published earlier \cite{Pri24,Zhang25Bristol}, indicating low background doping in the GaN channel with room temperature mobilities below $100\,\text{cm}^2\text{V}^{-1}\text{s}^{-1}$ and sheet carrier densities above $3.5\times 10^{13}\,\text{cm}^{-2}$.

For secondary ion mass spectroscopy (SIMS), we made a sample with an AlN-transition layer with double growth times and thicker (and hence relaxed) GaN with at the same growth conditions as for the devices as otherwise surface contamination would obscure the change of the thin layers, see Fig.~\ref{i:sample} c). Both carbon and oxygen drop quickly (within the $\approx 15$\,nm broadening of SIMS) to values below $10^{17}\text{cm}^{-3}$ inside the GaN. Still, the much higher values inside the AlN could also reduce the mobility, as discussed in another paper focusing on channel thickness \cite{Robin26} which contains more data on topography too. 

The simulation of bands and states were done with a Poisson/k*p couple solver (tibercad) with the parameters from ref.~\cite{Robin26}.

Device fabrication began with the deposition of a 5\,nm SiN using plasma-enhanced atomic layer deposition to protect the underlying N-polar layer during device processing. Source and drain contacts were defined by photolithography and then opened for the deposition of the ohmic metal stack of Ti/Al/Ni/Au on the GaN surface. After device isolation, T-shaped gates were defined by e-beam lithography. Subsequently, the Ni/Au gate contact was formed by e-beam evaporation. HEMT devices with different source-to-drain spacings ($L_{SD}$) and gate widths ($W_{G}$) were fabricated, while the gate length ($L_{G}$) of 250\,nm was unchanged. Finally, a SiN layer was deposited by plasma-enhanced chemical vapor deposition for surface passivation. After that, the contact-pad regions were opened for the following measurements. The final device cross-section is depicted in Fig.~\ref{i:static} a). The threshold voltage ($V_{th}$) was determined from a linear extrapolation of source-to-drain current ($I_{ds}$) at 6\,V source-drain bias to the X-axis intersection, as shown in Fig.~\ref{i:static} c).

\section{Results and Discussion}

\subsection{Electric Characteristics}

\begin{figure*}
\includegraphics[width=0.8\textwidth]{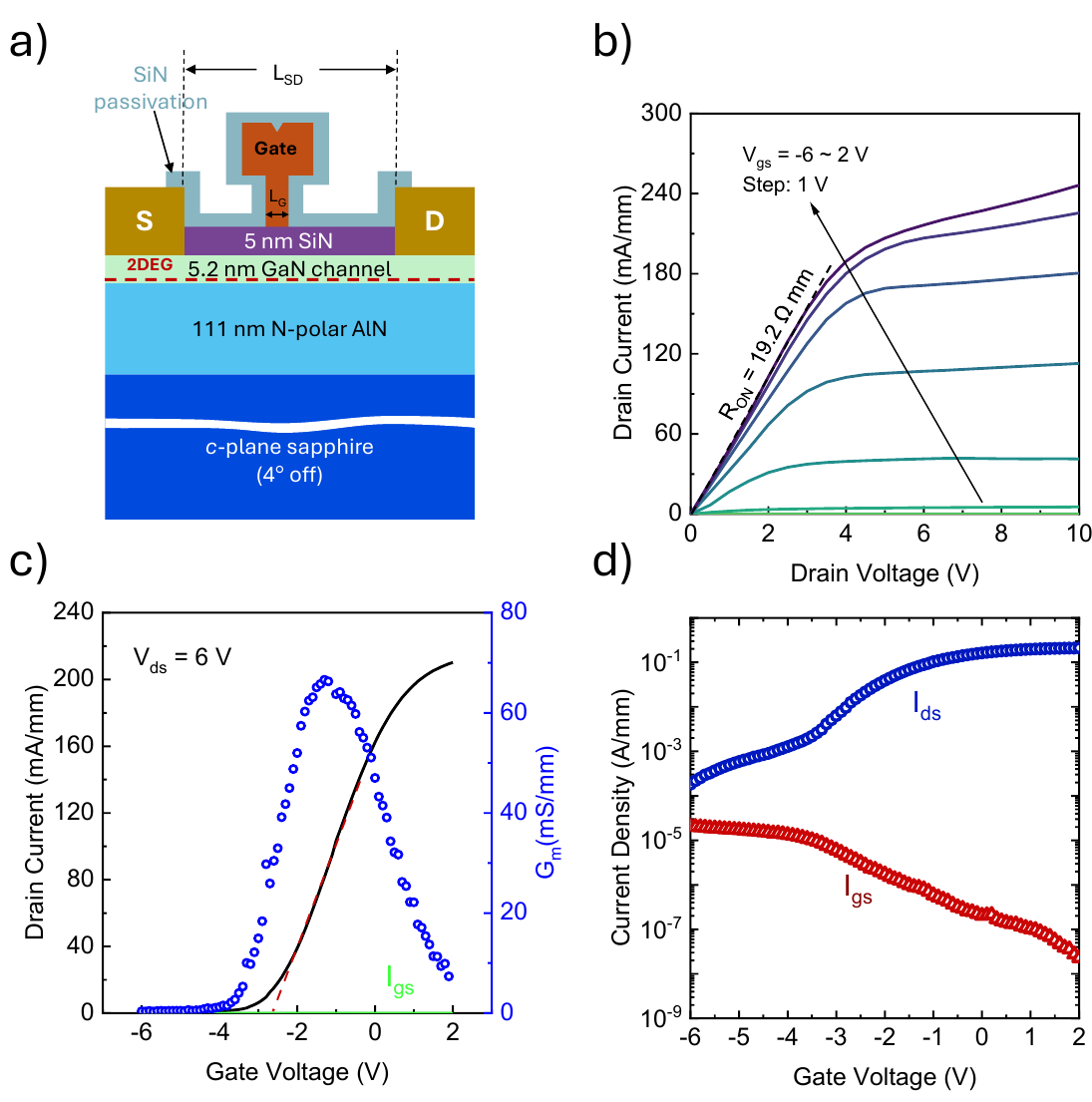}
\caption{(a) Cross-sectional representation of a processed HEMT. (b) Output characteristic for a 50\,µm $\times$ 2.4\,µm HEMT device. Gate-to-Source voltage $V_{GS}$ changed from -6 to 2\,V in the step of 1\,V. Corresponding linear (c) and semi-logarithmic (d) plots of transfer curves and gate leakage curves measured at $V_{ds}=6$\,V.}
\label{i:static}
\end{figure*}

The performance data of HEMTs with different dimensions are summarized in Table~\ref{t:all}. Figure ~\ref{i:static}b) shows the output characterisitics for a 50\,µm $\times$ 2.4\,µm ($W_{G}\times L_{SD}$) device, which exhibited an on resistance ($R_{ON}$) of 19.2\,$\Omega\,$mm (at 2\,V) and a $I_{sat}$ of approximately 0.24\,A/mm. Furthermore, the source-to-drain current ($I_{DS}$) from $V_{GS} = -4$ to $-6$\,V are all on top of the drain voltage ($V_{DS}$) axis, i.e. no current flow is visible in this linear plot. The good pinch-off behavior with off-state drain currents below 0.1\,mA/mm at $V_{GS} = -4$\,V can be seen more clearly in a semi-log plot. The off-state drain current ($< 10 \mu$A/mm at -6\,V) achieved in this work is over two orders of magnitude lower than the previously reported one for N-polar AlN-based HEMTs grown on bulk AlN substrates \cite{Kim23hemt}. Only our recent normally-off HEMT showed a higher on/off ratio but at the expense of a lower $I_{sat}$ \cite{Zhang25Bristol}.

Figures 2(c) and (d) show the transfer curves of this HEMT device. The threshold voltage ($V_{th}$) for this device was -2.7\,V from a linear extrapolation of $I_{DS}$ at 6,V drain bias to the X-axis intersection (Fig. 2 c). Figure 2(d) plots the source drain current ($I_{DS}$) at $V_{DS} = 6$\,V on a semi-log scale over the gate voltage, which yielded an on/off ratio (Ion/Ioff) of 1200 for this device. $V_{DS} = 6$\,V was chosen because it is close to $I_{sat}$. Using a lower $V_{DS}$ would of course reduce the on/off ratio because of a lower $I_{DS}$ in the on-state.

All the HEMTs in Table~\ref{t:all} demonstrated a good scaling behavior, where $R_{ON}$ decreased from 46.7 to 19.2\,$\Omega\,$mm and $I_{sat}$ increased from 0.13 to 0.24\,A/mm when shrinking the source-to-drain distance from 6 to 2.4\,µm. The largest $I_{sat}$ achieved in this study is close to the best devices on sapphire reported so far, with $I_{sat}=243$\,mA/mm (but at $V_g=3$\,V) and $R_{ON}\approx 23\,\Omega \,$mm (extracted from Fig.~3c in \cite{Kowaki24}). However, the devices have different dimensions and cannot be directly compared. Other reported $I_{sat}$ of N-polar AlN-based HEMTs on sapphire were considerably lower \cite{Inahara23,Miyamoto23,Zazuli24sims,Zazuli24thick} and hence had higher $R_{ON}$.

To date, the best N-polar AlN-based HEMTs have been realized by homo-epitaxy using MBE on AlN bulk substrates. As the properties scaled with dimensions, the most similar device is a 50\,µm $\times$ 4.0\,µm with a probably comparable process to ours (but thicker GaN channel), for which an $R_{ON}=4.12\,\Omega$ mm was reported with an $I_{sat}=1.2\,$A/mm (Fig.~2 b) in \cite{Kim23hemt}). These values scale approximately with the reported sheet resistance of $333\,\Omega/\Box$ versus our $2000\,\Omega/\Box$ in the center of our wafer. The much lower sheet resistance in the MBE-grown channels is due to a thicker channel and the significantly lower threading dislocation density due to homo-epitaxial growth as our current study of the GaN channel width suggests \cite{Robin26}. Furthermore, the lower oxygen content in the AlGaN layer directly below the GaN channel in the MBE channel may contribute, as the overlap of the 2DEG states with trap states in the back barrier is a critical issue \cite{Pri24,Robin26} and the impurities reported in N-polar AlN by MBE had been lower than ours (compare Fig.~\ref{i:sample}c) to Fig. 6 in \cite{Singhal22}). Without a dedicated AlN transition layer, we found a sheet resistance exceeding $10,000\,\Omega/\Box$. However, the optimization of our AlN transition layer to reduce the oxygen and carbon levels at the GaN/AlN interface is still ongoing.

The peak transconductance for our devices is 65\,mS/mm (Table~\ref{t:all}). Again, this is about $5\,\times$ lower than an aggressively scaled device on bulk N-polar AlN substrates, which reported 310\,mS/mm with a $L_{SD}=800$\,nm. A larger device on bulk AlN had 100\,mS/mm \cite{Kim22hemt}. For hetero-epitaxial N-polar AlN-based polarization-doped field effect transistors and HEMTs using AlGaN channels much lower peak transconductance were reported of 4.6\,mS/mm \cite{Lemettinen19} or even 3\,mS/mm \cite{Inahara23}.

The threshold voltages of our devices vary due to GaN channel thickness variations across the wafer since the latter strongly affect the sheet resistance \cite{Robin26}. They generally turn off between -2.9 and -2.4\,V (Tab.~\ref{t:all}). To date, no threshold voltages were given for N-polar AlN-based HEMTs with GaN channel in literature. However, from the published figures, one can extrapolate to a threshold voltage of about -6.5\,V \cite{Kim23hemt} for homo-epitaxy on AlN bulk, while the HEMTs with hetero-epitaxy on sapphire turn off around -3\,V \cite{Inahara23,Kowaki24,Zazuli24sims,Zazuli24thick}, similar to our devices. The only other reported hetero-epitaxial N-polar AlN-based field-effect transistor on SiC employed a much thicker polarisation-doped AlGaN channel \cite{Lemettinen19}, where this device turned off above -8\,V with a pinch-off at -18\,V \cite{Lemettinen19}.

\begin{table*}
\begin{tabular}{cc|ccccccc}
\hline
\hline
$W_G$&$L_{SD}$&$I_{sat}$&$V_{th}$&$G_m$&$I_{gs}$&$I_{on}/I_{off}$&$R_{on}$\\
(µm)&(µm)&(A/mm)&(V)&(mS/mm)&(A/mm)&ratio&($\Omega$\,mm)\\
\hline
37.5&2.4&0.17&-2.4&52&$1.5\times\,10^{-5}$&1300&28.4\\
50&2.4&0.24&-2.7&65&$2.0\times\,10^{-5}$&1200&19.2\\
50&4.0&0.21&-2.4&48&$2.3\times\,10^{-5}$&220&34.0\\
50&4.4&0.22&-2.9&50&$2.7\times\,10^{-5}$&900&29.6\\
50&5.0&0.18&-2.4&53&$1.8\times\,10^{-5}$&8700&33.8\\
50&5.5&0.15&-2.8&45&$2.6\times\,10^{-5}$&620&42.4\\
50&6.0&0.13&-2.7&30&$1.7\times\,10^{-5}$&770&46.7\\
75&2.4&0.24&-2.6&50&$3.1\times\,10^{-5}$&10100&21.4\\
100&2.4&0.25&-3.5&45&$3.3\times\,10^{-5}$&4200&20.2\\
\hline
\hline
\end{tabular}
\caption{Parameters for the HEMTs with different dimensions. $I_{sat}$ is measured at 2\,V and $I_{gs}$ at -6\,V.}
\label{t:all}
\end{table*}

\subsection{Benefit of the AlN/GaN interface}

To date, the N-polar AlN-based HEMTs in literature have usually very low on/off ratios below 100. For devices on bulk N-polar AlN substrates with SiN as the gate insulator, an on/off ratio of 12 was reported \cite{Kim22hemt}, and a more recent device using HfO$_2$-insulated gate still had only about 50 (extracted from Fig.~2d in \cite{Kim23hemt}). Inahara et al. reported an on/off ratio of around 70 for a HEMT on sapphire employing 10\% AlGaN channel \cite{Inahara23}. There are no publications for hetero-epitaxial N-polar AlN-based HEMTs with GaN channels, but a pronounced off-state leakage current can be seen in the best published data (Fig.~4 in \cite{Kowaki24} and \cite{Zazuli24thick}). A SiC-based N-polar AlGaN polarisation-doped field-effect transistor reported a high on/off ratio of 11000, but required a voltage swing of 20\,V \cite{Lemettinen19}. Even N-polar GaN-based HEMTs often have small on/off ratios and thus are seldom reported, mostly just seen indirectly from transfer characteristics plotted on a linear scale (e.g. several figures in \cite{Singisetti2013review}). A high on/off ratio of 2.2 $\times$ $10^5$ was achieved for an enhanced mode HEMT, where the 2DEG under the gate was compensated by a 2D hole gas \cite{Singisetti2013review}. Very recently, similar on/off ratios of $\approx$\,$10^5$ have also been reported by two depletion-mode N-polar GaN-based HEMTs with AlGaN capping layers. \cite{Akso24,Soman25}.

In contrast, our recently published N-polar AlN-based enhanced-mode HEMT using the same epitaxial structure as the devices in this paper reached even an on/off ratio of $10^7$ (at the expense of a lower $I_{sat}$) \cite{Zhang25Bristol}. Devices in this work have on/off ratios between 620 and 10,000, most often around 1000. All the N-polar AlN-based HEMTs discussed above have an AlGaN layer below the GaN channel, while our epitaxial structure did not include such an AlGaN layer, but employ a direct transition from AlN to GaN.

\begin{figure}
\includegraphics[width=1.0\columnwidth]{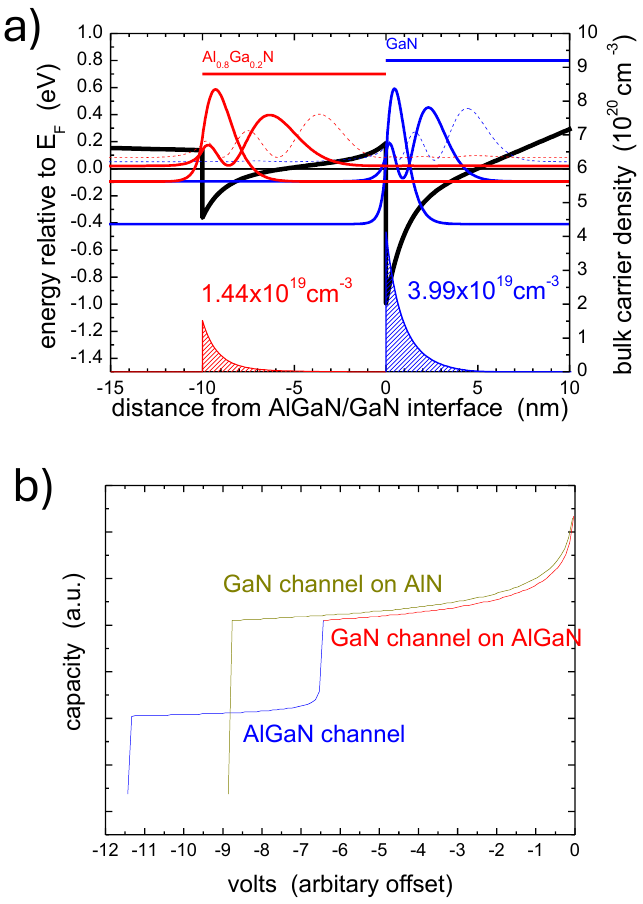}
\caption{a) band diagram (left axis) with the states of the AlN/AlGaN 2DEG (red) and the AlGaN/GaN 2DEG (blue). Right axis is the sheet electron concentration. b) calculated capacitance from a) indications two thresholds for an AlGaN intermediate layer.}
\label{i:sim}
\end{figure}

Simulation reported in \cite{Kim23hemt} found that the AlGaN layer may generate a 2DEG at the AlGaN/AlN interface. This is confirmed by our own simulations in Fig.~\ref{i:sim}. Even though 80\,\% of AlGaN was assumed for the 10\,nm interlayer below the GaN channel, this already reduces the 2DEG carrier concentrations in the GaN channel by a quarter. Moreover, the ground state of the AlGaN 2DEG and the 1st excited state of the GaN 2DEG matches closely in energy and could interact; the 2nd excited states (only $\approx 0.12$\,eV above the Fermi level) of both have even a visible non-zero probability in their overlap region, making it possible to couple the two 2DEG. And even though the AlGaN 2DEG may not contribute much to conductivity (due to its low mobility with alloy scattering and the carriers having to contact it through the AlGaN), the second 2DEG in AlGaN would hold up the Fermi-level like a buried gate since it is difficult to deplete by a top gate.
Next, let’s calculate the expected voltages to deplete the second AlGaN 2DEG. The formula would be very simple for constant carrier densities. But our carrier densities are not constant. Hence, we calculate a CV curve instead. The capacity is given by
\begin{equation}
    C(z)=8.85\times 10^7 A \varepsilon_r \frac1d
\end{equation}
with $A$ the area in cm$^2$ and $z$ the depletion depth from the surface in nm. The voltage is then given by
\begin{equation}
V(z)=\frac{\varepsilon_r A^2}{1.41\times 10^{32}}\int^z_0\frac{\partial n}{\partial C^2}dx+V_0
\end{equation}
with $n$ carrier density at depth $z$ (i.e. from simulation) and $V_0$ a random offset voltage from integration. Since the capacity $C$ depends on $A$, the area is cancelled out for the final voltage.

Numerically, the integration was done in Octave/Matlab and shown in Fig.~\ref{i:sim} b). The main channel carrier density (as expressed by the capacity) can be strongly reduced by applying -3\,V, very close to the threshold found in Fig.~\ref{i:static} and reported for other devices \cite{Kowaki24,Inahara23,Zazuli24sims,Zazuli24thick}. However, to deplete the second AlGaN channel, another -7\,V are needed. This extra voltage depends strongly on the thickness and Al content, so small variations in the structure could alter this voltage. Nevertheless, it is clear from Fig.~\ref{i:sim} b) that depleting the second 2DEG requires considerable effort or, if not depleted, the second AlGaN 2DEG could act as a leakage path similar to buffer leakage in conventional GaN-based HEMTs \cite{Kotani23,Yu10leak}. The situation is comparable to N-polar GaN-based HEMTs, where the carriers have to cross the AlGaN back barrier and the buffer leakage occurs in the N-polar GaN below.

Thus, our choice of the abrupt GaN/AlN heterostructure which circumvent this potential issue. This comes with a price, however, the 2DEG states can overlap with traps in the first 1-2 nm of the AlN back barrier \cite{Robin26}. Hence, one has to reduce the oxygen and carbon contents in the N-polar AlN directly below the GaN channel. For this, we employ ramping of the NH$_3$ flow and temperature in our AlN transition layer as seen in Fig.~\ref{i:sample} c). To shorten the growth interruption from high-temperature AlN buffer growth to low-temperature GaN channel growth steps, we ramped down the temperature during the transition layer. To not increase background carbon in the AlN too much when decreasing the growth temperature, we also increased the NH$_3$ flow correspondingly. Increasing NH$_3$ flow indeed lowered the carbon incorporation but first increased the oxygen background, probably due to residual water in the NH$_3$. However, the oxygen decreased when the temperature was lowered. This might be due to more hydrogen from partly decomposed NHx staying longer on the surface at lower temperature, but further investigation is needed to confirm this.

Although the difference in carbon and oxygen concentrations between the AlN transition layer and the AlN buffer layer seems small on the large logarithmic scale, the inclusion of the transition layer growth at lower temperature allows for shortening the growth interruption from 7\,minutes, to 50\,seconds, and by this effectively eliminating the interfacial impurity spike between the GaN channel and the AlN buffer observed in \cite{Pri24}.
Overall, the carbon and especially the oxygen impurity levels are a constant challenges for N-polar AlN and GaN and further optimization is ongoing.

\subsection{Gate Leakage}

Apart from the abrupt GaN/AlN structure, the improved gate leakage that is particularly critical for surface-sensitive N-polar devices as the channel is very close to the gate and without an intermediate barrier layer. Significant gate leakage establishes an unwanted current path, which substantially increases the off-state current and degrades the on/off ratio \cite{Kim22hemt}. As shown in Fig. \ref{i:static} d) and Tab.~\ref{t:all}, our devices have typical gate leakage currents of $1.5-3\times 10^{-5}$\,A/mm. This is much lower than for the first N-polar AlN-based HEMT on bulk AlN with SiN as the gate dielectric as in our structures, for which a much higher gate leakage close to $10^{-1}$\,A/mm was reported \cite{Kim22hemt}. A N-polar polarization-doped field effect transistor on SiC with AlN below the gate had a comparable gate leakage to our structures, but the channel was almost 30\,nm thick and thus had itself a high resistance \cite{Lemettinen19}. More comparable to our devices are N-polar GaN-based HEMTs which also use thin GaN channels. For instance, with HfO$_2$ gate dielectric also a higher gate leakage of $10^{-2}$\,A/mm was reported \cite{Meyer11HfO}. Recently reported N-polar GaN-based HEMTs exhibited lower gate leakage of $10^{-5}$\,A/mm \cite{Akso24,Soman25}, comparable to our devices under similar gate bias conditions.

One key issue is the smoothness of our surfaces, which RMS roughnesses below 0.5\,nm for both interface and GaN channel surface (see Fig.~\ref{i:sample} a) inset), due to a rather weak step-bunching background \cite{Robin26}. Together with a high-quality gate dielectric, this suppresses gate leakage.

Unlike the two-terminal measurement, where we found the minimum gate leakage to be near 0\,V gate voltage, the three-terminal transfer curve in Fig.~\ref{i:static} d) shows a decreasing leakage current as the gate voltage increases from -6 to 2\,V. This discrepancy is likely caused by the influence of the drain-source bias present during transfer characterization. The applied $V_{DS}$ in transfer characteristic measurements could alter the electric field distribution and tilts the barrier, producing asymmetric leakage paths to the source and drain and leading to gate leakage current minimum potentially moves positive. Moreover, the channel potential beneath the gate could be lifted locally under a $V_{DS}$ bias, therefore modifying the effective gate-to-channel electric field and contributing to the shift of gate leakage minimum toward positive gate voltage. These effects discussed above serve as possible causes to explain the behavior observed in Fig.~\ref{i:static} d).

\subsection{Switching Behavior}

The subthreshold slope in our HEMT is around 1V/decade (Fig.~\ref{i:static} d), while the hysteresis from the bidirectional gate sweep is about 200\,mV (data not shown). Both are indicators of weak gate control and likely originate from a high density of interface states at the SiN/N-polar GaN interface, maybe in combination with the high threading dislocation density above $10^10\,\text{cm}^2{-2}$. The issues above are the primary factors currently limiting further improvements in on/off ratio and leakage current and represent key focuses for our future optimization.

\begin{figure}
\includegraphics[width=1.0\columnwidth]{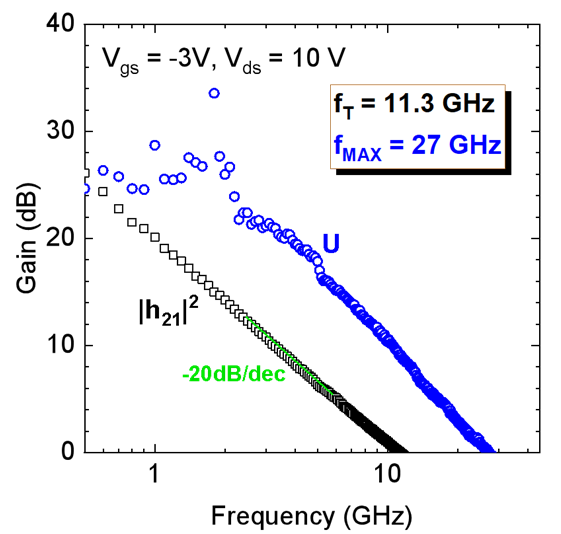}
\caption{Measured cutoff frequencies of a HEMT with $L_G = 250\,$nm in this study, showing highest $f_T=11.3$\,GHz and $f_{MAX} = 27\,$GHz.}
\label{i:RF}
\end{figure}

Overall, the difference in on-state performance between our hetero-epitaxial N-polar AlN-based HEMT on sapphire and the ones reported on bulk AlN substrates scale with about 6 times lower sheet resistance of the latter. Consequently, our RF properties of a $50\,\mu m \times  2.4\,mu m$ device have the same issue, showing the current gain cut-off frequency $f_T \approx 11.3$\,GHz and the maximum oscillation frequency $f_{MAX}\approx 27$\,GHz (Fig.~\ref{i:RF}). Furthermore, the high oxygen concentration (on the order of $10^{19}\,\text{cm}^{-3}$) and the dislocation density in our AlN could result in deep pinning states near the AlN/GaN interface, impeding high frequency performance and, as mentioned, are another focus of optimization.

\section{Conclusion}

We have realized N-polar AlN-based GaN channel HEMTs grown on sapphire substrates by MOVPE. These devices show high on/off ratios of 1000 and higher, which we attribute to the abrupt AlN/GaN interface between buffer and channel and low buffer leakage. The HEMT performance is currently limited by the sheet resistance of the GaN channel due to dislocations from the growth on sapphire, residual oxygen and carbon impurities, and probably also by the interface states between GaN channel and the gate dielectric. Further progress will address all these points.

\section*{acknowledgment}

We thank Dr. Yoann Robin for helpful discussion. This work was partly supported JSPS KAKENHI (Grant No. JPJSJRP 20221603) and JST SPRING, Japan Grant Number JPMJSP2125.

\bibliography{eigene,N-polar,HEMT,AlN}% Produces the bibliography via BibTeX.

@article{Yoshikawa24,
	doi = {10.35848/1347-4065/ad565a},
	year = {2024},
	month = {jun},
	publisher = {IOP Publishing},
	volume = {63},
	number = {6},
	pages = {060903},
	author = {Yoshikawa, Akira and Nagatomi, Takaharu and Nagase, Kazuhiro and Sugiyama, Sho and Schowalter, Leo J.},
	title = {Pseudomorphic growth of a thin-{GaN} layer on the {AlN} single-crystal substrate using metal organic vapor phase epitaxy},
	journal = {Jpn. J. Appl. Phys.}
}

@article{Meyer11HfO,
    author = {Meyer, D. J. and Katzer, D. S. and Deen, D. A. and Storm, D. F. and Binari, S. C. and Gougousi, T.},
    title = {HfO$_2$-insulated gate {N-polar GaN HEMTs} with high breakdown voltage},
    journal = {physica status solidi (a)},
    volume = {208},
    number = {7},
    pages = {1630-1633},
    doi = {10.1002/pssa.201001080},
    year = {2011}
}

@article{Singisetti2013review,
	doi = {10.1088/0268-1242/28/7/074006},
	year = {2013},
	month = {jun},
	publisher = {IOP Publishing},
	volume = {28},
	number = {7},
	pages = {074006},
	author = {Singisetti, Uttam and Wong, Man Hoi and Mishra, Umesh K},
	title = {High-performance {N-polar GaN} enhancement-mode device technology},
	journal = {Semicond. Sci. Technol.}
}

@article{Soman25,
    doi = {10.35848/1882-0786/adcb87},
    year = {2025},
    month = {apr},
    publisher = {IOP Publishing},
    volume = {18},
    number = {4},
    pages = {046503},
    author = {Soman, Rohith and Malakoutian, Mohamadali and Kim, Jeong-kyu and Akso, Emre and Hatui, Nirupam and Wurm, Christian and Mishra, Umesh and Chowdhury, Srabanti},
    title = {Integration of 150\,nm gate length {N-polar GaN MIS-HEMT} devices with all-around diamond for device-level cooling},
    journal = {APEX}
}

@ARTICLE{Akso24,
	author={Akso, Emre and Collins, Henry and Khan, Kamruzzaman and Wang, Boyu and Li, Weiyi and Clymore, Christopher and Kayede, Emmanuel and Liu, Wenjian and Chavan, Tanmay and Hamwey, Robert and Hatui, Nirupam and Guidry, Matthew and Romanczyk, Brian and Keller, Stacia and Mishra, Umesh K.},
	journal={IEEE Microwave and Wireless Technology Letters}, 
	title={Schottky Barrier Gate {N-Polar GaN}-on-Sapphire Deep Recess {HEMT} With Record {10.5\,dB} Linear Gain and 50.2\,\% {PAE} at {94\,GHz}}, 
	year={2024},
	volume={34},
	number={2},
	pages={183-186},
	doi={10.1109/LMWT.2023.3345531}
}

@article{Kotani23,
	author = {Kotani, Junji and Makiyama, Kozo and Ohki, Toshihiro and Ozaki, Shiro and Okamoto, Naoya and Minoura, Yuichi and Sato, Masaru and Nakamura, Norikazu and Miyamoto, Yasuyuki},
	title = {High-power-density {InAlGaN/GaN HEMT} using {InGaN} back barrier for {W}-band amplifiers},
	journal = {Electronics Letters},
	volume = {59},
	number = {4},
	pages = {e12715},
	doi = {10.1049/ell2.12715},
	year = {2023}
}

@article{Yu10leak,
    author = {Cao, Yu and Zimmermann, Tom and Xing, Huili and Jena, Debdeep},
    title = {Polarization-engineered removal of buffer leakage for {GaN} transistors},
    journal = {Applied Physics Letters},
    volume = {96},
    number = {4},
    pages = {042102},
    year = {2010},
    month = {01},
    doi = {10.1063/1.3293454},
}

@article{Hu23AlN,
    author = {Hu, Mingtao and Wang, Ping and Wang, Ding and Wu, Yuanpeng and Mondal, Shubham and Wang, Danhao and Ahmadi, Elaheh and Ma, Tao and Mi, Zetian},
    title = {Heteroepitaxy of N-polar AlN on C-face 4H-SiC: Structural and optical properties},
    journal = {APL Materials},
    volume = {11},
    number = {12},
    pages = {121111},
    year = {2023},
    month = {12},
    issn = {2166-532X},
    doi = {10.1063/5.0168970},
	key= "Rought N-AlN"
}

@article{Isono20,
	author = {Isono, Tatsuya and Ito, Tadatoshi and Sakamoto, Ryota and Yao, Yongzhao and Ishikawa, Yukari and Okada, Narihito and Tadatomo, Kazuyuki},
	title = "Growth of {N}-Polar Aluminum Nitride on Vicinal Sapphire Substrates and Aluminum Nitride Bulk Substrates",
	journal = {phys. stat. solidi (b)},
	volume = {257},
	number = {4},
	pages = {1900588},
	doi = {10.1002/pssb.201900588},
	year = {2020}
}

@article{Miyamoto23,
	doi = {10.35848/1347-4065/acf8cf},
	year = {2023},
	month = {oct},
	publisher = {IOP Publishing},
	volume = {62},
	number = {SN},
	pages = {SN1016},
	author = {Minagi Miyamoto and Wataru Matsumura and Ryo Okuno and Syunsuke Matsuda and Koki Hanasaku and Taketo Kowaki and Daisuke Inahara and Satoshi Kurai and Narihito Okada and Yoichi Yamada},
	title = {Improvement of electrical properties by insertion of {AlGaN} interlayer for {N-polar AlGaN/AlN} structures on sapphire substrates},
	journal = {Jpn. J. Appl. Phys.}
}

@article{Inahara23,
	author = {Inahara, Daisuke and Matsuda, Shunsuke and Matsumura, Wataru and Okuno, Ryo and Hanasaku, Koki and Kowaki, Taketo and Miyamoto, Minagi and Yao, Yongzhao and Ishikawa, Yukari and Tanaka, Atsushi and Honda, Yoshio and Nitta, Shugo and Amano, Hiroshi and Kurai, Satoshi and Okada, Narihito and Yamada, Yoichi},
	title = {Investigation of Electrical Properties of {N}-Polar {AlGaN/AlN} Heterostructure Field-Effect Transistors},
	journal = {phys. stat. solidi (a)},
	volume = {220},
	number = {16},
	pages = {2200871},
	doi = {10.1002/pssa.202200871},
	year = {2023}
}

@article{Kowaki24,
	author = {Kowaki, Taketo and Hanasaku, Koki and Miyamoto, Minagi and Zazuli, Aina Hiyama and Inahara, Daisuke and Fujii, Kai and Kimoto, Taisei and Ninoki, Ryosuke and Kurai, Satoshi and Okada, Narihito and Yamada, Yoichi},
	title = {Effect of the Twist Crystallinity of {N}-Polar {AlN} Underlayer on the Electrical Properties of {GaN/AlN} Structures},
	journal = {physica status solidi (a)},
	volume = {221},
	number = {21},
	pages = {2400053},
    year = {2024},
	doi = {10.1002/pssa.202400053}
}

@article{Zazuli24sims,
	author = {Zazuli, Aina Hiyama and Kowaki, Taketo and Miyamoto, Minagi and Hanasaku, Koki and Inahara, Daisuke and Fujii, Kai and Kurai, Satoshi and Okada, Narihito and Yamada, Yoichi},
	title = {Electrical Properties of {N}-Polar {GaN/AlGaN/AlN} Grown via Metal-Organic Vapor Phase Epitaxy},
	journal = {physica status solidi (a)},
	volume = {221},
    number = {21},
	pages = {2400060},
	year={2024},
	doi = {10.1002/pssa.202400060}
}

@article{Zazuli24thick,
	doi = {10.35848/1347-4065/ad6e8f},
	year = {2024},
	month = {sep},
	publisher = {IOP Publishing},
	volume = {63},
	number = {9},
	pages = {09SP11},
	author = {Zazuli, Aina Hiyama and Kowaki, Taketo and Miyamoto, Minagi and Hanasaku, Koki and Inahara, Daisuke and Fujii, Kai and Kimoto, Taisei and Ninoki, Ryosuke and Kurai, Satoshi and Okada, Narihito and Yamada, Yoichi},
	title = {Impact of thick {N}-polar {AlN} growth on crystalline quality and electrical properties of {N}-polar {GaN/AlGaN/AlN} {FET}},
	journal = {Jpn. J. Appl. Phys.}
}

@article{Singhal22,
    author = {Singhal, Jashan and Encomendero, Jimy and Cho, Yongjin and van Deurzen, Len and Zhang, Zexuan and Nomoto, Kazuki and Toita, Masato and Xing, Huili Grace and Jena, Debdeep},
    title = {Molecular beam homoepitaxy of {N-polar AlN} on bulk {AlN} substrates},
    journal = {AIP Advances},
    volume = {12},
    number = {9},
    pages = {095314},
    year = {2022},
    month = {09},
    doi = {10.1063/5.0100225},
	key="MBE growth with SIMS"
}

@article{Lemettinen18sims,
	doi = {10.7567/APEX.11.101002},
	year = {2018},
	month = {sep},
	publisher = {The Japan Society of Applied Physics},
	volume = {11},
	number = {10},
	pages = {101002},
	author = {Lemettinen, Jori and Okumura, Hironori and Palacios, Tomás and Suihkonen, Sami},
	title = {N-polar {AlN} buffer growth by metal–organic vapor phase epitaxy for transistor applications},
	journal = {Appl. Phys. Express}
}

@ARTICLE{Lemettinen19,
    author={Lemettinen, Jori and Chowdhury, Nadim and Okumura, Hironori and Kim, Iurii and Suihkonen, Sami and Palacios, Tomás},
    journal={IEEE Electron Device Letters}, 
    title={Nitrogen-Polar Polarization-Doped Field-Effect Transistor Based on {Al$_{0.8}$Ga$_{0.2}$N/AlN} on {SiC} With Drain Current Over 100\,mA/mm}, 
    year={2019},
    volume={40},
    number={8},
    pages={1245-1248},
    doi={10.1109/LED.2019.2923902},
    key="N-polar HFET by polarisation doping"
}

@article{Namikawa23,
	title = {MOVPE growth of AlN and AlGaN films on N-polar annealed and sputtered AlN templates},
	journal = {J. Crystal Growth},
	volume = {617},
	pages = {127256},
	year = {2023},
	issn = {0022-0248},
	doi = {10.1016/j.jcrysgro.2023.127256},
	author = {Gaku Namikawa and Kanako Shojiki and Riku Yoshida and Ryusei Kusuda and Kenjiro Uesugi and Hideto Miyake}
}

@inproceedings{Kim22hemt,
	author={Kim, Eungkyun and Zhang, Zexuan and Singhal, Jashan and Nomoto, Kazuki and Hickman, Austin and Toita, Masato and Jena, Debdeep and Xing, Huili Grace},
	booktitle={2022 Device Research Conference (DRC)}, 
	title={First demonstration of {N}-polar {GaN/AlGaN/AlN HEMT} on Single Crystal {AlN} Substrates}, 
	year={2022},
	pages={1-2},
	doi={10.1109/DRC55272.2022.9855776},
	key="First N-polar HEMT"
}

@article{Kim23hemt,
    author = {Kim, Eungkyun and Zhang, Zexuan and Encomendero, Jimy and Singhal, Jashan and Nomoto, Kazuki and Hickman, Austin and Wang, Cheng and Fay, Patrick and Toita, Masato and Jena, Debdeep and Xing, Huili Grace},
    title = "{N-polar GaN/AlGaN/AlN} high electron mobility transistors on single-crystal bulk {AlN} substrates",
    journal = {Appl. Phys. Lett.},
    volume = {122},
    number = {9},
    pages = {092104},
    year = {2023},
    month = {03},
    issn = {0003-6951},
    doi = {10.1063/5.0138939},
}

@Article{Yang25dislocations,
	author ="Yang, Yuming and Zhang, Xuemei and Qin, Mi and Liu, Jun and Zhang, Chuanguo and Hui, Zhixin and Li, Yonggang and Zeng, Zhi and Zhang, Yongsheng",
	title  ="The effect of interface polarity on the basal dislocations at the {GaN/AlN} interface",
	journal  ="Phys. Chem. Chem. Phys.",
	year  ="2025",
	volume  ="27",
	issue  ="1",
	pages  ="355-366",
	publisher  ="The Royal Society of Chemistry",
	doi  ="10.1039/D4CP03069A"
}

@Article{Yang25SiCepi,
	AUTHOR = {Yang, Yong and Ni, Xianfeng and Fan, Qian and Gu, Xing},
	TITLE = {Study of Low-Temperature {(Al)GaN} on {N-Polar GaN} Films Grown by {MOCVD} on Vicinal {SiC} Substrates},
	JOURNAL = {Materials},
	VOLUME = {18},
	YEAR = {2025},
	NUMBER = {3},
	ARTICLE-NUMBER = {638},
	DOI = {10.3390/ma18030638}
}

@article{Pampili24,
    author = {Pampili, Pietro and Pristovsek, Markus},
    title = "Nitrogen-polar growth of {AlN} on vicinal (0001) sapphire by {MOVPE}",
    journal = {J. Appl. Phys.},
    volume = {135},
    number = {19},
    pages = {195303},
    year = {2024},
    month = {05},
    issn = {0021-8979},
    doi = {10.1063/5.0202746},
}

@Article{Pri24,
	AUTHOR = {Pristovsek, Markus and Furuhashi, Itsuki and Yang, Xu and Zhang, Chengzhi and Smith, Matthew D.},
	TITLE = {Two-Dimensional Electron Gas in Thin {N-}Polar {GaN} Channels on {AlN} on Sapphire Templates},
	JOURNAL = {Crystals},
	VOLUME = {14},
	YEAR = {2024},
	NUMBER = {9},
	ARTICLE-NUMBER = {822},
	URL = {https://www.mdpi.com/2073-4352/14/9/822},
	ISSN = {2073-4352},
	DOI = {10.3390/cryst14090822}
}

@article{Zhang25Bristol,
	author={Zhang, Chengzhi and Yin, Yidi and Huang, Peng and Furuhashi, Itsuki and Yoann, Robin and Pristovsek, Markus and Kuball, Martin and Smith, Matthew David},
	title={N-polar {AlN-}based enhancement-mode transistor with {p-NiO$_x$} gate stacks and reduced buffer trapping},
	journal = "J. Phys. D: Appl. Phys.",
	doi={10.1088/1361-6463/ae161c},
	year={2025}
}

@article{Robin26,
	author={Robin Yoann and Itsuki Furuhashi and Markus Pristovsek},
	title={The limits of electrical transport in thin {GaN} channels on {N-polar AlN}},
	journal = "J. Semiconductors",
	year={2026},
	volume={in print},
	doi={10.1088/1674-4926/26010034}
}

\end{document}